\newcounter{@sc}
\newcounter{@scp}
\newcounter{@t}
\newlength{\@x}
\newlength{\@xa}
\newlength{\@xb}
\newlength{\@y}
\newlength{\@ya}
\newlength{\@yb}
\newsavebox{\@pt}

\def\bezier#1(#2,#3)(#4,#5)(#6,#7){\c@@sc#1\relax
  \c@@scp\c@@sc \advance\c@@scp\@ne
  \@xb #4\unitlength \advance\@xb -#2\unitlength \multiply\@xb \tw@
  \@xa #6\unitlength \advance\@xa -#2\unitlength
      \advance\@xa -\@xb \divide\@xa\c@@sc
  \@yb #5\unitlength \advance\@yb -#3\unitlength \multiply\@yb \tw@
  \@ya #7\unitlength \advance\@ya -#3\unitlength
      \advance\@ya -\@yb \divide\@ya\c@@sc
  \setbox\@pt\hbox{\vrule height\@halfwidth  depth\@halfwidth
   width\@wholewidth}\c@@t\z@
   \put(#2,#3){\@whilenum{\c@@t<\c@@scp}\do
      {\@x\c@@t\@xa \advance\@x\@xb \divide\@x\c@@sc \multiply\@x\c@@t
       \@y\c@@t\@ya \advance\@y\@yb \divide\@y\c@@sc \multiply\@y\c@@t
       \raise \@y \hbox to \z@{\hskip \@x\unhcopy\@pt\hss}%
       \advance\c@@t\@ne}}}

\endinput

\documentstyle[bezier]{article}
\font\tenbf=cmbx10
\font\tenrm=cmr10
\font\tenit=cmti10
\font\elevenbf=cmbx10 scaled\magstep 1
\font\elevenrm=cmr10 scaled\magstep 1
\font\elevenit=cmti10 scaled\magstep 1

\renewenvironment{thebibliography}[1]
 { \elevenrm
   \begin{list}{\arabic{enumi}.}
    {\usecounter{enumi} \setlength{\parsep}{0pt}
     \setlength{\itemsep}{3pt} \settowidth{\labelwidth}{#1.}
     \sloppy
    }}{\end{list}}

\newcommand{\densityw}{20}

\newcommand{\be}{\begin{equation}}
\newcommand{\ee}{\end{equation}}
\newcommand{\bea}{\begin{eqnarray}}
\newcommand{\eea}{\end{eqnarray}}
\newcommand{\ep}{\varepsilon}
\newcommand{\Int}{\int\limits}

\newcommand{\lnyx}{\ln \frac{y}{x}}

\begin{document}

\vspace{-1cm}

\begin{center}{{\tenbf PROGRESS IN EVALUATING SOME COMPLICATED TYPES\\
               \vglue 3pt
               OF FEYNMAN DIAGRAMS WITH TWO (AND MORE) LOOPS \\
               \vglue 5pt
{To appear in Proceedings of the AIHENP-93 workshop \\
(Oberammergau, Germany, 4--8 October 1993)}
\\}
\vglue 5pt
\vglue 1.0cm
{\tenrm A. I. DAVYDYCHEV \\}
\baselineskip=13pt
{\tenit Institute for Nuclear Physics, Moscow State University,\\}
\baselineskip=12pt
{\tenit 119899, Moscow, Russia \\}
{\tenrm and\\}
\baselineskip=13pt
{\tenit Department of Physics, University of Bergen, \\ }
\baselineskip=12pt
{\tenit All\'egaten 55, N-5007 Bergen, Norway\\}
}
\end{center}
\vglue 0.5cm
{\rightskip=3pc
 \leftskip=3pc
 \tenrm\baselineskip=12pt
 \noindent
   Problems occurring in physically important non-trivial
examples of loop calculations are discussed.
   A procedure of deriving expansions of two-loop self-energy
diagrams with different masses is constructed. The cases
of small and large external momentum are considered.
The coefficients of the expansions are calculated analytically.
Comparison with numerical results shows good agreement
below the first threshold and above the last threshold
of the diagram.
   An approach to evaluating three- and four-point ladder
diagrams with massless internal particles and off-shell
external momenta is presented. Exact results for an arbitrary
number of loops are obtained. Evaluating non-planar diagrams
is also discussed.
\vglue 0.6cm}
\baselineskip=14pt
\elevenrm
In the present talk I would like to discuss some recent results for
loop diagrams obtained in collaboration with J.B.~Tausk
$^{1,2}$,
V.A.~Smirnov $^{2}$
and N.I.~Ussyukina $^{3,4,5}$.
The papers $^{1,2}$
are devoted to the
examination of massive two-loop self-energy diagrams, whilst
refs.~$^{3,4,5}$
deal with some three- and four-point
diagrams with massless internal particles and off-shell
external momenta.
\vglue 0.6cm
{\elevenbf\noindent 1. Two-loop massive self-energy diagrams}
\vglue 0.4cm
It is known that any two-loop self-energy diagram (see Fig.~1)
\begin{figure}[b]
\setlength{\unitlength}{0.2mm}
\raisebox{-61.2\unitlength}{
\put(20.0,0.0){
\begin{picture}(314.3,190)( 0.0,-120)
\put( 0.0, 0.0){\line(1,0){96.0}}
\bezier{\densityw}(96.0, 0.0)(96.0,-4.0)(96.5,-8.0)
\bezier{\densityw}(96.5,-8.0)(97.0,-12.0)(98.1,-15.8)
\bezier{\densityw}(98.1,-15.8)(99.1,-19.7)(100.7,-23.4)
\bezier{\densityw}(100.7,-23.4)(102.2,-27.1)(104.2,-30.6)
\bezier{\densityw}(104.2,-30.6)(106.2,-34.1)(108.6,-37.2)
\bezier{\densityw}(108.6,-37.2)(111.1,-40.4)(113.9,-43.2)
\bezier{\densityw}(113.9,-43.2)(116.7,-46.1)(119.9,-48.5)
\bezier{\densityw}(119.9,-48.5)(123.1,-51.0)(126.6,-53.0)
\bezier{\densityw}(126.6,-53.0)(130.1,-55.0)(133.8,-56.5)
\bezier{\densityw}(133.8,-56.5)(137.5,-58.0)(141.3,-59.1)
\bezier{\densityw}(141.3,-59.1)(145.2,-60.1)(149.2,-60.6)
\bezier{\densityw}(149.2,-60.6)(153.2,-61.2)(157.2,-61.2)
\bezier{\densityw}(157.2,-61.2)(161.2,-61.2)(165.1,-60.6)
\bezier{\densityw}(165.1,-60.6)(169.1,-60.1)(173.0,-59.1)
\bezier{\densityw}(173.0,-59.1)(176.9,-58.0)(180.6,-56.5)
\bezier{\densityw}(180.6,-56.5)(184.3,-55.0)(187.7,-53.0)
\bezier{\densityw}(187.7,-53.0)(191.2,-51.0)(194.4,-48.5)
\bezier{\densityw}(194.4,-48.5)(197.6,-46.1)(200.4,-43.2)
\bezier{\densityw}(200.4,-43.2)(203.2,-40.4)(205.7,-37.2)
\bezier{\densityw}(205.7,-37.2)(208.1,-34.1)(210.1,-30.6)
\bezier{\densityw}(210.1,-30.6)(212.1,-27.1)(213.7,-23.4)
\bezier{\densityw}(213.7,-23.4)(215.2,-19.7)(216.2,-15.8)
\bezier{\densityw}(216.2,-15.8)(217.3,-12.0)(217.8,-8.0)
\bezier{\densityw}(217.8,-8.0)(218.3,-4.0)(218.3, 0.0)
\bezier{\densityw}(218.3, 0.0)(218.3, 4.0)(217.8, 8.0)
\bezier{\densityw}(217.8, 8.0)(217.3,12.0)(216.2,15.8)
\bezier{\densityw}(216.2,15.8)(215.2,19.7)(213.7,23.4)
\bezier{\densityw}(213.7,23.4)(212.1,27.1)(210.1,30.6)
\bezier{\densityw}(210.1,30.6)(208.1,34.1)(205.7,37.2)
\bezier{\densityw}(205.7,37.2)(203.2,40.4)(200.4,43.2)
\bezier{\densityw}(200.4,43.2)(197.6,46.1)(194.4,48.5)
\bezier{\densityw}(194.4,48.5)(191.2,51.0)(187.7,53.0)
\bezier{\densityw}(187.7,53.0)(184.3,55.0)(180.6,56.5)
\bezier{\densityw}(180.6,56.5)(176.9,58.0)(173.0,59.1)
\bezier{\densityw}(173.0,59.1)(169.1,60.1)(165.1,60.6)
\bezier{\densityw}(165.1,60.6)(161.2,61.2)(157.2,61.2)
\put(218.3, 0.0){\line(1,0){96.0}}
\bezier{\densityw}(157.2,61.2)(153.2,61.2)(149.2,60.6)
\bezier{\densityw}(149.2,60.6)(145.2,60.1)(141.3,59.1)
\bezier{\densityw}(141.3,59.1)(137.5,58.0)(133.8,56.5)
\bezier{\densityw}(133.8,56.5)(130.1,55.0)(126.6,53.0)
\bezier{\densityw}(126.6,53.0)(123.1,51.0)(119.9,48.5)
\bezier{\densityw}(119.9,48.5)(116.7,46.1)(113.9,43.2)
\bezier{\densityw}(113.9,43.2)(111.1,40.4)(108.6,37.2)
\bezier{\densityw}(108.6,37.2)(106.2,34.1)(104.2,30.6)
\bezier{\densityw}(104.2,30.6)(102.2,27.1)(100.7,23.4)
\bezier{\densityw}(100.7,23.4)(99.1,19.7)(98.1,15.8)
\bezier{\densityw}(98.1,15.8)(97.0,12.0)(96.5, 8.0)
\bezier{\densityw}(96.5, 8.0)(96.0, 4.0)(96.0, 0.0)
\put(157.2,61.2){\line(0,-1){122.3}}
\put (   20,  7){\vector(1, 0){40}}
\put (254.3,  7){\vector(1, 0){40}}
\put (   40, 14){\makebox(0,0)[bl]{\large $k$}}
\put (274.3, 14){\makebox(0,0)[bl]{\large $k$}}
\put (  122, 35){\makebox(0,0)[c]{\large $4$}}
\put (  122,-35){\makebox(0,0)[c]{\large $1$}}
\put (192.5, 35){\makebox(0,0)[c]{\large $5$}}
\put (192.5,-35){\makebox(0,0)[c]{\large $2$}}
\put (  168,  0){\makebox(0,0)[c]{\large $3$}}
\put (157,-94){\makebox(0,0)[c]{\large ${\mbox{(a)}}$}}
\end{picture}
}}
\raisebox{-61.2\unitlength}{
\put(400.0,0.0){
\begin{picture}(314.3,190)( 0.0,-120)
\put( 0.0, 0.0){\line(1,0){96.0}}
\bezier{\densityw}(96.0, 0.0)(96.0,-4.0)(96.5,-8.0)
\bezier{\densityw}(96.5,-8.0)(97.0,-12.0)(98.1,-15.8)
\bezier{\densityw}(98.1,-15.8)(99.1,-19.7)(100.7,-23.4)
\bezier{\densityw}(100.7,-23.4)(102.2,-27.1)(104.2,-30.6)
\bezier{\densityw}(104.2,-30.6)(106.2,-34.1)(108.6,-37.2)
\bezier{\densityw}(108.6,-37.2)(111.1,-40.4)(113.9,-43.2)
\bezier{\densityw}(113.9,-43.2)(116.7,-46.1)(119.9,-48.5)
\bezier{\densityw}(119.9,-48.5)(123.1,-51.0)(126.6,-53.0)
\bezier{\densityw}(126.6,-53.0)(130.1,-55.0)(133.8,-56.5)
\bezier{\densityw}(133.8,-56.5)(137.5,-58.0)(141.3,-59.1)
\bezier{\densityw}(141.3,-59.1)(145.2,-60.1)(149.2,-60.6)
\bezier{\densityw}(149.2,-60.6)(153.2,-61.2)(157.2,-61.2)
\bezier{\densityw}(157.2,-61.2)(161.2,-61.2)(165.1,-60.6)
\bezier{\densityw}(165.1,-60.6)(169.1,-60.1)(173.0,-59.1)
\bezier{\densityw}(173.0,-59.1)(176.9,-58.0)(180.6,-56.5)
\bezier{\densityw}(180.6,-56.5)(184.3,-55.0)(187.7,-53.0)
\bezier{\densityw}(187.7,-53.0)(191.2,-51.0)(194.4,-48.5)
\bezier{\densityw}(194.4,-48.5)(197.6,-46.1)(200.4,-43.2)
\bezier{\densityw}(200.4,-43.2)(203.2,-40.4)(205.7,-37.2)
\bezier{\densityw}(205.7,-37.2)(208.1,-34.1)(210.1,-30.6)
\bezier{\densityw}(210.1,-30.6)(212.1,-27.1)(213.7,-23.4)
\bezier{\densityw}(213.7,-23.4)(215.2,-19.7)(216.2,-15.8)
\bezier{\densityw}(216.2,-15.8)(217.3,-12.0)(217.8,-8.0)
\bezier{\densityw}(217.8,-8.0)(218.3,-4.0)(218.3, 0.0)
\put(218.3, 0.0){\line(1,0){96.0}}
\bezier{\densityw}(96.0, 0.0)(96.0, 4.0)(96.5, 8.0)
\bezier{\densityw}(96.5, 8.0)(97.0,12.0)(98.1,15.8)
\bezier{\densityw}(98.1,15.8)(99.1,19.7)(100.7,23.4)
\bezier{\densityw}(100.7,23.4)(102.2,27.1)(104.2,30.6)
\bezier{\densityw}(104.2,30.6)(106.2,34.1)(108.6,37.2)
\bezier{\densityw}(218.3, 0.0)(218.3, 4.0)(217.8, 8.0)
\bezier{\densityw}(217.8, 8.0)(217.3,12.0)(216.2,15.8)
\bezier{\densityw}(216.2,15.8)(215.2,19.7)(213.7,23.4)
\bezier{\densityw}(213.7,23.4)(212.1,27.1)(210.1,30.6)
\bezier{\densityw}(210.1,30.6)(208.1,34.1)(205.7,37.2)
\bezier{\densityw}(108.6,37.2)(110.9,34.3)(113.5,31.6)
\bezier{\densityw}(113.5,31.6)(116.2,28.9)(119.1,26.6)
\bezier{\densityw}(119.1,26.6)(122.0,24.3)(125.2,22.3)
\bezier{\densityw}(125.2,22.3)(128.4,20.4)(131.8,18.8)
\bezier{\densityw}(131.8,18.8)(135.2,17.3)(138.8,16.1)
\bezier{\densityw}(138.8,16.1)(142.3,15.0)(146.0,14.3)
\bezier{\densityw}(146.0,14.3)(149.7,13.6)(153.4,13.4)
\bezier{\densityw}(153.4,13.4)(157.2,13.2)(160.9,13.4)
\bezier{\densityw}(160.9,13.4)(164.6,13.6)(168.3,14.3)
\bezier{\densityw}(168.3,14.3)(172.0,15.0)(175.6,16.1)
\bezier{\densityw}(175.6,16.1)(179.1,17.3)(182.5,18.8)
\bezier{\densityw}(182.5,18.8)(185.9,20.4)(189.1,22.3)
\bezier{\densityw}(189.1,22.3)(192.3,24.3)(195.2,26.6)
\bezier{\densityw}(195.2,26.6)(198.2,28.9)(200.8,31.6)
\bezier{\densityw}(200.8,31.6)(203.4,34.3)(205.7,37.2)
\bezier{\densityw}(108.6,37.2)(110.9,40.2)(113.5,42.9)
\bezier{\densityw}(113.5,42.9)(116.2,45.5)(119.1,47.9)
\bezier{\densityw}(119.1,47.9)(122.0,50.2)(125.2,52.1)
\bezier{\densityw}(125.2,52.1)(128.4,54.1)(131.8,55.7)
\bezier{\densityw}(131.8,55.7)(135.2,57.2)(138.8,58.3)
\bezier{\densityw}(138.8,58.3)(142.3,59.5)(146.0,60.1)
\bezier{\densityw}(146.0,60.1)(149.7,60.8)(153.4,61.0)
\bezier{\densityw}(153.4,61.0)(157.2,61.3)(160.9,61.0)
\bezier{\densityw}(160.9,61.0)(164.6,60.8)(168.3,60.1)
\bezier{\densityw}(168.3,60.1)(172.0,59.5)(175.6,58.3)
\bezier{\densityw}(175.6,58.3)(179.1,57.2)(182.5,55.7)
\bezier{\densityw}(182.5,55.7)(185.9,54.1)(189.1,52.1)
\bezier{\densityw}(189.1,52.1)(192.3,50.2)(195.2,47.9)
\bezier{\densityw}(195.2,47.9)(198.2,45.5)(200.8,42.9)
\bezier{\densityw}(200.8,42.9)(203.4,40.2)(205.7,37.2)
\put (   20,  7){\vector(1, 0){40}}
\put (254.3,  7){\vector(1, 0){40}}
\put (   40, 14){\makebox(0,0)[bl]{\large $k$}}
\put (274.3, 14){\makebox(0,0)[bl]{\large $k$}}
\put (  108, 16){\makebox(0,0)[c]{\large $1$}}
\put (157.2, 50){\makebox(0,0)[c]{\large $2$}}
\put (157.2, 24){\makebox(0,0)[c]{\large $3$}}
\put (206.3, 16){\makebox(0,0)[c]{\large $4$}}
\put (157.2,-50){\makebox(0,0)[c]{\large $5$}}
\put (157,-94){\makebox(0,0)[c]{\large ${\mbox{(b)}}$}}
\end{picture}
}}
\caption{Two-loop self-energy diagrams}
\end{figure}
can be reduced to scalar integrals $^{6}$.
Moreover, in the case
when the powers of denominators are integer, the diagram in Fig.~1b
can be reduced to Fig.~1a by the decomposition of the first and the
fourth denominators. So, we can express all two-loop two-point
contributions in terms of the scalar integrals corresponding to
Fig.~1a,
\be
\label{defJ}
 J(\{\nu_i\} ; \{m_i\} ; k) = \int \int
\frac{\mbox{d}^n p \; \mbox{d}^n q}
     {D_1^{\nu_1} D_2^{\nu_2} D_3^{\nu_3} D_4^{\nu_4} D_5^{\nu_5} } ,
\ee
where $D_i = (p_i^2-m_i^2+i0)$ are massive denominators,
$\nu_i$ are
the powers of these denominators, $n=4-2\ep$ is the space-time
dimension $^{7}$,
and $p_i$ are constructed from the external momentum
$k$ and the loop integration momenta $p$ and $q$ (with due account of the
momentum conservation).

For some special cases the integrals (\ref{defJ}) can be evaluated
exactly $^{8,9}$
and expessed in terms of
the polylogarithms $\mbox{Li}_N$ $^{10}$
with $N \leq 3$.
On the other hand, the problem of evaluating such diagrams when all the
internal lines are massive is more complicated, and exact expressions
are not known. For example, in ref.~$^{11}$
the result was
presented in terms of a two-fold integral representation.

In the papers $^{1,2}$
we studied expansions of massive two-loop
two-point diagrams for the case of small $^{1}$
and large $^{2}$
values of the external momentum squared. Note that in general
the diagram in Fig.~1a has four physical thresholds (with respect
to $k^2$), corresponding to different cuts of the diagram; namely,
\[
(m_1+m_4)^2 , \;\; (m_2+m_5)^2 , \;\; (m_1+m_3+m_5)^2, \;\;
(m_2+m_3+m_4)^2 .
\]
The cases considered correspond
to the situations when we are either below the lowest threshold~$^{1}$
or above the highest one~$^{2}$.

For small values of $k^2$, the expansion of the integral (\ref{defJ})
is a usual Taylor expansion (we consider the general case, when the
lowest thershold is not equal to zero). Coefficients of the expansion
can be represented in terms of two-loop massive ``vacuum" diagrams.
Moreover, for the case when all powers of denominators are integer,
they can be reduced to vacuum diagrams, each of them depending no more
than on three different masses (see Fig.~2).
\begin{figure}[htb]
\centering
\setlength{\unitlength}{0.250mm}
\raisebox{-61.2\unitlength}{
\begin{picture}(314.3,122.3)( 0.0,-61.2)
\bezier{\densityw}(96.0, 0.0)(96.0,-4.0)(96.5,-8.0)
\bezier{\densityw}(96.5,-8.0)(97.0,-12.0)(98.1,-15.8)
\bezier{\densityw}(98.1,-15.8)(99.1,-19.7)(100.7,-23.4)
\bezier{\densityw}(100.7,-23.4)(102.2,-27.1)(104.2,-30.6)
\bezier{\densityw}(104.2,-30.6)(106.2,-34.1)(108.6,-37.2)
\bezier{\densityw}(108.6,-37.2)(111.1,-40.4)(113.9,-43.2)
\bezier{\densityw}(113.9,-43.2)(116.7,-46.1)(119.9,-48.5)
\bezier{\densityw}(119.9,-48.5)(123.1,-51.0)(126.6,-53.0)
\bezier{\densityw}(126.6,-53.0)(130.1,-55.0)(133.8,-56.5)
\bezier{\densityw}(133.8,-56.5)(137.5,-58.0)(141.3,-59.1)
\bezier{\densityw}(141.3,-59.1)(145.2,-60.1)(149.2,-60.6)
\bezier{\densityw}(149.2,-60.6)(153.2,-61.2)(157.2,-61.2)
\bezier{\densityw}(157.2,-61.2)(161.2,-61.2)(165.1,-60.6)
\bezier{\densityw}(165.1,-60.6)(169.1,-60.1)(173.0,-59.1)
\bezier{\densityw}(173.0,-59.1)(176.9,-58.0)(180.6,-56.5)
\bezier{\densityw}(180.6,-56.5)(184.3,-55.0)(187.7,-53.0)
\bezier{\densityw}(187.7,-53.0)(191.2,-51.0)(194.4,-48.5)
\bezier{\densityw}(194.4,-48.5)(197.6,-46.1)(200.4,-43.2)
\bezier{\densityw}(200.4,-43.2)(203.2,-40.4)(205.7,-37.2)
\bezier{\densityw}(205.7,-37.2)(208.1,-34.1)(210.1,-30.6)
\bezier{\densityw}(210.1,-30.6)(212.1,-27.1)(213.7,-23.4)
\bezier{\densityw}(213.7,-23.4)(215.2,-19.7)(216.2,-15.8)
\bezier{\densityw}(216.2,-15.8)(217.3,-12.0)(217.8,-8.0)
\bezier{\densityw}(217.8,-8.0)(218.3,-4.0)(218.3, 0.0)
\bezier{\densityw}(218.3, 0.0)(218.3, 4.0)(217.8, 8.0)
\bezier{\densityw}(217.8, 8.0)(217.3,12.0)(216.2,15.8)
\bezier{\densityw}(216.2,15.8)(215.2,19.7)(213.7,23.4)
\bezier{\densityw}(213.7,23.4)(212.1,27.1)(210.1,30.6)
\bezier{\densityw}(210.1,30.6)(208.1,34.1)(205.7,37.2)
\bezier{\densityw}(205.7,37.2)(203.2,40.4)(200.4,43.2)
\bezier{\densityw}(200.4,43.2)(197.6,46.1)(194.4,48.5)
\bezier{\densityw}(194.4,48.5)(191.2,51.0)(187.7,53.0)
\bezier{\densityw}(187.7,53.0)(184.3,55.0)(180.6,56.5)
\bezier{\densityw}(180.6,56.5)(176.9,58.0)(173.0,59.1)
\bezier{\densityw}(173.0,59.1)(169.1,60.1)(165.1,60.6)
\bezier{\densityw}(165.1,60.6)(161.2,61.2)(157.2,61.2)
\bezier{\densityw}(157.2,61.2)(153.2,61.2)(149.2,60.6)
\bezier{\densityw}(149.2,60.6)(145.2,60.1)(141.3,59.1)
\bezier{\densityw}(141.3,59.1)(137.5,58.0)(133.8,56.5)
\bezier{\densityw}(133.8,56.5)(130.1,55.0)(126.6,53.0)
\bezier{\densityw}(126.6,53.0)(123.1,51.0)(119.9,48.5)
\bezier{\densityw}(119.9,48.5)(116.7,46.1)(113.9,43.2)
\bezier{\densityw}(113.9,43.2)(111.1,40.4)(108.6,37.2)
\bezier{\densityw}(108.6,37.2)(106.2,34.1)(104.2,30.6)
\bezier{\densityw}(104.2,30.6)(102.2,27.1)(100.7,23.4)
\bezier{\densityw}(100.7,23.4)(99.1,19.7)(98.1,15.8)
\bezier{\densityw}(98.1,15.8)(97.0,12.0)(96.5, 8.0)
\bezier{\densityw}(96.5, 8.0)(96.0, 4.0)(96.0, 0.0)
\put(157.2,61.2){\line(0,-1){122.3}}
\put (  150, 20){\vector(0,-1){40}}
\put (   92, 37){\vector(1, 1){28}}
\put (194.5, 65){\vector(1,-1){28}}
\put (  143,  0){\makebox(0,0)[r]{\large $p-q$}}
\put (   98, 59){\makebox(0,0)[r]{\large $p$}}
\put (216.5, 59){\makebox(0,0)[l]{\large $q$}}
\put (  122,-35){\makebox(0,0)[c]{\large $1$}}
\put (192.5,-35){\makebox(0,0)[c]{\large $2$}}
\put (  168,  0){\makebox(0,0)[c]{\large $3$}}
\end{picture}
}
\caption{The vacuum two-loop diagram}

\end{figure}
These diagrams with higher (integer) powers of denominators
(occurring in the coefficients of the expansion) can be evaluated
by use of the integration-by-parts technique $^{12}$ (see also $^{13}$),
and all of them (up to finite in $\ep$ parts) can be
represented in terms of dilogarithms or Clausen's function $^{10}$.
By use of the {\sf REDUCE} system $^{14}$,
we constructed an algorithm of analytical
evaluation of the coefficients of the expansion.
We compared the results for some diagrams with a numerical
calculation based on the integral representation~$^{11}$,
and we found a nice agreement in the region below the lowest threshold.

The case when $k^2$ is larger than the highest threshold of the diagram
is more
complicated, because the corresponding expansion is not a usual Taylor
expansion, but also contains logarithms and squared logarithms of
$-k^2$ (in four dimensions) yielding an imaginary part when the
momentum is time-like. To obtain this expansion, we applied
a general mathematical theorem on asymptotic expansions of Feynman
integrals in the limit of large external momentum.
Expansions of this kind were presented in refs.~$^{15,16}$.
For our case (\ref{defJ}), the asymptotic expansion theorem gives
\be
\label{theorem}
J_{\Gamma} \begin{array}{c} \frac{}{}  \\
                    {\mbox{\Large$\sim$}} \\ {}_{k^2 \to \infty}
            \end{array}
\sum_{\gamma} J_{\Gamma/\gamma} \; \circ  \;
{\cal{T}}_{\{m_i\}; \{q_i\}} J_{\gamma} ,
\ee
where $\Gamma$ is the main graph (see Fig.1a), $\gamma$ are subgraphs involved
in the asymptotic expansion (see below), $\Gamma/\gamma$ is the reduced graph
obtained from $\Gamma$ by shrinking the subgraph $\gamma$ to a single point,
$J_{\gamma}$ denotes
the dimensionally-regularized Feynman integral corresponding to a graph
$\gamma$ (for example, $J_{\Gamma}$ is given by (\ref{defJ})),
${\cal{T}}_{\{m_i\}; \{q_i\}}$ is the operator of Taylor expansion of the
integrand in masses and ``small'' momenta $q_i$ (that are ``external''
for the given subgraph $\gamma$, but do not contain the ``large'' external
momentum $k$), and the symbol ``$\circ$'' means that the resulting
polynomial in these momenta should be inserted into the numerator of
the integrand of $J_{\Gamma/\gamma}$. It is implied that the
operator $\cal{T}$ acts on the integrands before the loop integrations
are performed.

In our case (see Fig.~1a), the sum (\ref{theorem}) goes over all
subgraphs $\gamma$ that become one-particle irreducible when we
connect the two vertices with external momentum $k$ by a line.
These subgraphs (there are five different types of
them) are shown in Fig.~3 (dotted lines correspond to the lines
that do not belong to $\gamma$).
\newcommand{\suba}{
\setlength{\unitlength}{0.253mm}
\raisebox{-15.5\unitlength}{
\begin{picture}(78.9,30.9)( 0.0,-15.5)
\put( 0.0, 0.0){\line(1,0){24.0}}
\bezier{\densityw}(24.0, 0.0)(24.0,-4.1)(26.1,-7.7)
\bezier{\densityw}(26.1,-7.7)(28.1,-11.3)(31.7,-13.4)
\bezier{\densityw}(31.7,-13.4)(35.3,-15.5)(39.5,-15.5)
\bezier{\densityw}(39.5,-15.5)(43.6,-15.5)(47.2,-13.4)
\bezier{\densityw}(47.2,-13.4)(50.8,-11.3)(52.8,-7.7)
\bezier{\densityw}(52.8,-7.7)(54.9,-4.1)(54.9, 0.0)
\bezier{\densityw}(54.9, 0.0)(54.9, 4.1)(52.8, 7.7)
\bezier{\densityw}(52.8, 7.7)(50.8,11.3)(47.2,13.4)
\bezier{\densityw}(47.2,13.4)(43.6,15.5)(39.5,15.5)
\put(54.9, 0.0){\line(1,0){24.0}}
\bezier{\densityw}(39.5,15.5)(35.3,15.5)(31.7,13.4)
\bezier{\densityw}(31.7,13.4)(28.1,11.3)(26.1, 7.7)
\bezier{\densityw}(26.1, 7.7)(24.0, 4.1)(24.0, 0.0)
\put(39.5,15.5){\line(0,-1){30.9}}
\end{picture}
} }
\newcommand{\subb}{
\setlength{\unitlength}{0.069mm}
\raisebox{-56.1\unitlength}{
\begin{picture}(288.2,112.1)( 0.0,-56.1)
\put( 0.0, 0.0){\line(1,0){88.0}}
\bezier{\densityw}(88.0, 0.0)(88.0,-0.4)(88.0,-0.9)
\bezier{\densityw}(88.7,-8.8)(88.8,-9.2)(88.8,-9.6)
\bezier{\densityw}(90.7,-17.3)(90.9,-17.7)(91.0,-18.2)
\bezier{\densityw}(94.1,-25.5)(94.3,-25.9)(94.5,-26.2)
\bezier{\densityw}(98.7,-33.0)(99.0,-33.3)(99.2,-33.7)
\bezier{\densityw}(104.4,-39.7)(104.7,-40.0)(105.1,-40.3)
\bezier{\densityw}(111.1,-45.4)(111.5,-45.6)(111.8,-45.9)
\bezier{\densityw}(118.6,-50.0)(119.0,-50.2)(119.4,-50.4)
\bezier{\densityw}(126.8,-53.3)(127.2,-53.5)(127.6,-53.6)
\bezier{\densityw}(135.3,-55.4)(135.7,-55.5)(136.2,-55.5)
\bezier{\densityw}(144.1,-56.1)(148.1,-56.1)(152.1,-55.5)
\bezier{\densityw}(152.1,-55.5)(156.0,-54.9)(159.9,-53.8)
\bezier{\densityw}(159.9,-53.8)(163.7,-52.7)(167.4,-51.0)
\bezier{\densityw}(167.4,-51.0)(171.0,-49.3)(174.4,-47.2)
\bezier{\densityw}(174.4,-47.2)(177.8,-45.0)(180.8,-42.4)
\bezier{\densityw}(180.8,-42.4)(183.8,-39.8)(186.5,-36.7)
\bezier{\densityw}(186.5,-36.7)(189.1,-33.7)(191.2,-30.3)
\bezier{\densityw}(191.2,-30.3)(193.4,-27.0)(195.1,-23.3)
\bezier{\densityw}(195.1,-23.3)(196.7,-19.7)(197.9,-15.8)
\bezier{\densityw}(197.9,-15.8)(199.0,-12.0)(199.6,-8.0)
\bezier{\densityw}(199.6,-8.0)(200.2,-4.0)(200.2,-0.0)
\bezier{\densityw}(200.2,-0.0)(200.2, 4.0)(199.6, 8.0)
\bezier{\densityw}(199.6, 8.0)(199.0,11.9)(197.9,15.8)
\bezier{\densityw}(197.9,15.8)(196.7,19.6)(195.1,23.3)
\bezier{\densityw}(195.1,23.3)(193.4,26.9)(191.2,30.3)
\bezier{\densityw}(191.2,30.3)(189.1,33.7)(186.5,36.7)
\bezier{\densityw}(186.5,36.7)(183.8,39.7)(180.8,42.4)
\bezier{\densityw}(180.8,42.4)(177.8,45.0)(174.4,47.2)
\bezier{\densityw}(174.4,47.2)(171.0,49.3)(167.4,51.0)
\bezier{\densityw}(167.4,51.0)(163.7,52.7)(159.9,53.8)
\bezier{\densityw}(159.9,53.8)(156.0,54.9)(152.1,55.5)
\bezier{\densityw}(152.1,55.5)(148.1,56.1)(144.1,56.1)
\put(200.2,-0.0){\line(1,0){88.0}}
\bezier{\densityw}(144.1,56.1)(140.1,56.1)(136.1,55.5)
\bezier{\densityw}(136.1,55.5)(132.1,54.9)(128.3,53.8)
\bezier{\densityw}(128.3,53.8)(124.4,52.7)(120.8,51.0)
\bezier{\densityw}(120.8,51.0)(117.1,49.3)(113.8,47.2)
\bezier{\densityw}(113.8,47.2)(110.4,45.0)(107.4,42.4)
\bezier{\densityw}(107.4,42.4)(104.3,39.7)(101.7,36.7)
\bezier{\densityw}(101.7,36.7)(99.1,33.7)(96.9,30.3)
\bezier{\densityw}(96.9,30.3)(94.7,26.9)(93.1,23.3)
\bezier{\densityw}(93.1,23.3)(91.4,19.6)(90.3,15.8)
\bezier{\densityw}(90.3,15.8)(89.2,11.9)(88.6, 8.0)
\bezier{\densityw}(88.6, 8.0)(88.0, 4.0)(88.0,-0.0)
\put(144.1,56.1){\line(0,-1){112.1}}
\end{picture}
} }
\newcommand{\subc}{
\setlength{\unitlength}{0.069mm}
\raisebox{-56.1\unitlength}{
\begin{picture}(288.2,112.2)( 0.0,-56.1)
\put( 0.0, 0.0){\line(1,0){88.0}}
\bezier{\densityw}(88.0, 0.0)(88.0,-4.0)(88.6,-8.0)
\bezier{\densityw}(88.6,-8.0)(89.1,-11.9)(90.3,-15.8)
\bezier{\densityw}(90.3,-15.8)(91.4,-19.6)(93.1,-23.3)
\bezier{\densityw}(93.1,-23.3)(94.7,-26.9)(96.9,-30.3)
\bezier{\densityw}(96.9,-30.3)(99.1,-33.7)(101.7,-36.7)
\bezier{\densityw}(101.7,-36.7)(104.3,-39.7)(107.4,-42.4)
\bezier{\densityw}(107.4,-42.4)(110.4,-45.0)(113.8,-47.2)
\bezier{\densityw}(113.8,-47.2)(117.1,-49.3)(120.8,-51.0)
\bezier{\densityw}(120.8,-51.0)(124.4,-52.7)(128.3,-53.8)
\bezier{\densityw}(128.3,-53.8)(132.1,-54.9)(136.1,-55.5)
\bezier{\densityw}(136.1,-55.5)(140.1,-56.1)(144.1,-56.1)
\bezier{\densityw}(144.1,-56.1)(144.5,-56.1)(145.0,-56.1)
\bezier{\densityw}(152.8,-55.4)(153.3,-55.3)(153.7,-55.2)
\bezier{\densityw}(161.4,-53.3)(161.8,-53.2)(162.2,-53.0)
\bezier{\densityw}(169.5,-50.0)(169.9,-49.8)(170.3,-49.6)
\bezier{\densityw}(177.0,-45.4)(177.4,-45.1)(177.7,-44.8)
\bezier{\densityw}(183.7,-39.6)(184.0,-39.3)(184.3,-39.0)
\bezier{\densityw}(189.4,-33.0)(189.7,-32.6)(190.0,-32.2)
\bezier{\densityw}(194.0,-25.4)(194.2,-25.1)(194.4,-24.7)
\bezier{\densityw}(197.4,-17.3)(197.5,-16.9)(197.7,-16.5)
\bezier{\densityw}(199.5,-8.8)(199.5,-8.3)(199.6,-7.9)
\bezier{\densityw}(200.2, 0.0)(200.2, 4.0)(199.6, 8.0)
\bezier{\densityw}(199.6, 8.0)(199.0,12.0)(197.9,15.8)
\bezier{\densityw}(197.9,15.8)(196.7,19.7)(195.1,23.3)
\bezier{\densityw}(195.1,23.3)(193.4,27.0)(191.2,30.3)
\bezier{\densityw}(191.2,30.3)(189.1,33.7)(186.5,36.7)
\bezier{\densityw}(186.5,36.7)(183.8,39.8)(180.8,42.4)
\bezier{\densityw}(180.8,42.4)(177.8,45.0)(174.4,47.2)
\bezier{\densityw}(174.4,47.2)(171.0,49.3)(167.4,51.0)
\bezier{\densityw}(167.4,51.0)(163.7,52.7)(159.9,53.8)
\bezier{\densityw}(159.9,53.8)(156.0,54.9)(152.1,55.5)
\bezier{\densityw}(152.1,55.5)(148.1,56.1)(144.1,56.1)
\put(200.2, 0.0){\line(1,0){88.0}}
\bezier{\densityw}(144.1,56.1)(140.1,56.1)(136.1,55.5)
\bezier{\densityw}(136.1,55.5)(132.1,54.9)(128.3,53.8)
\bezier{\densityw}(128.3,53.8)(124.4,52.7)(120.8,51.0)
\bezier{\densityw}(120.8,51.0)(117.1,49.3)(113.8,47.2)
\bezier{\densityw}(113.8,47.2)(110.4,45.0)(107.4,42.4)
\bezier{\densityw}(107.4,42.4)(104.3,39.8)(101.7,36.7)
\bezier{\densityw}(101.7,36.7)(99.1,33.7)(96.9,30.3)
\bezier{\densityw}(96.9,30.3)(94.7,27.0)(93.1,23.3)
\bezier{\densityw}(93.1,23.3)(91.4,19.7)(90.3,15.8)
\bezier{\densityw}(90.3,15.8)(89.2,12.0)(88.6, 8.0)
\bezier{\densityw}(88.6, 8.0)(88.0, 4.0)(88.0, 0.0)
\put(144.1,56.1){\line(0,-1){112.2}}
\end{picture}
} }
\newcommand{\subd}{
\setlength{\unitlength}{0.069mm}
\raisebox{-56.1\unitlength}{
\begin{picture}(288.1,112.1)( 0.0,-56.1)
\put( 0.0, 0.0){\line(1,0){88.0}}
\bezier{\densityw}(88.0, 0.0)(88.0,-4.0)(88.6,-8.0)
\bezier{\densityw}(88.6,-8.0)(89.1,-11.9)(90.3,-15.8)
\bezier{\densityw}(90.3,-15.8)(91.4,-19.6)(93.1,-23.3)
\bezier{\densityw}(93.1,-23.3)(94.7,-26.9)(96.9,-30.3)
\bezier{\densityw}(96.9,-30.3)(99.1,-33.7)(101.7,-36.7)
\bezier{\densityw}(101.7,-36.7)(104.3,-39.7)(107.4,-42.4)
\bezier{\densityw}(107.4,-42.4)(110.4,-45.0)(113.8,-47.2)
\bezier{\densityw}(113.8,-47.2)(117.1,-49.3)(120.8,-51.0)
\bezier{\densityw}(120.8,-51.0)(124.4,-52.7)(128.3,-53.8)
\bezier{\densityw}(128.3,-53.8)(132.1,-54.9)(136.1,-55.5)
\bezier{\densityw}(136.1,-55.5)(140.1,-56.1)(144.1,-56.1)
\bezier{\densityw}(144.1,-56.1)(148.1,-56.1)(152.0,-55.5)
\bezier{\densityw}(152.0,-55.5)(156.0,-54.9)(159.9,-53.8)
\bezier{\densityw}(159.9,-53.8)(163.7,-52.7)(167.4,-51.0)
\bezier{\densityw}(167.4,-51.0)(171.0,-49.3)(174.4,-47.2)
\bezier{\densityw}(174.4,-47.2)(177.8,-45.0)(180.8,-42.4)
\bezier{\densityw}(180.8,-42.4)(183.8,-39.7)(186.4,-36.7)
\bezier{\densityw}(186.4,-36.7)(189.1,-33.7)(191.2,-30.3)
\bezier{\densityw}(191.2,-30.3)(193.4,-26.9)(195.1,-23.3)
\bezier{\densityw}(195.1,-23.3)(196.7,-19.6)(197.9,-15.8)
\bezier{\densityw}(197.9,-15.8)(199.0,-11.9)(199.6,-8.0)
\bezier{\densityw}(199.6,-8.0)(200.1,-4.0)(200.1, 0.0)
\bezier{\densityw}(200.1, 0.0)(200.1, 4.0)(199.6, 8.0)
\bezier{\densityw}(199.6, 8.0)(199.0,11.9)(197.9,15.8)
\bezier{\densityw}(197.9,15.8)(196.7,19.6)(195.1,23.3)
\bezier{\densityw}(195.1,23.3)(193.4,26.9)(191.2,30.3)
\bezier{\densityw}(191.2,30.3)(189.1,33.7)(186.4,36.7)
\bezier{\densityw}(186.4,36.7)(183.8,39.7)(180.8,42.4)
\bezier{\densityw}(180.8,42.4)(177.8,45.0)(174.4,47.2)
\bezier{\densityw}(174.4,47.2)(171.0,49.3)(167.4,51.0)
\bezier{\densityw}(167.4,51.0)(163.7,52.7)(159.9,53.8)
\bezier{\densityw}(159.9,53.8)(156.0,54.9)(152.0,55.5)
\bezier{\densityw}(152.0,55.5)(148.1,56.1)(144.1,56.1)
\put(200.1, 0.0){\line(1,0){88.0}}
\bezier{\densityw}(144.1,56.1)(143.6,56.1)(143.2,56.1)
\bezier{\densityw}(135.3,55.4)(134.9,55.3)(134.4,55.2)
\bezier{\densityw}(126.7,53.3)(126.3,53.2)(125.9,53.0)
\bezier{\densityw}(118.6,50.0)(118.2,49.8)(117.8,49.6)
\bezier{\densityw}(111.1,45.4)(110.8,45.1)(110.4,44.8)
\bezier{\densityw}(104.4,39.6)(104.1,39.3)(103.8,39.0)
\bezier{\densityw}(98.7,33.0)(98.4,32.6)(98.2,32.2)
\bezier{\densityw}(94.1,25.4)(93.9,25.1)(93.7,24.7)
\bezier{\densityw}(90.7,17.3)(90.6,16.9)(90.5,16.5)
\bezier{\densityw}(88.7, 8.8)(88.6, 8.3)(88.5, 7.9)
\put(144.1,56.1){\line(0,-1){112.1}}
\end{picture}
} }
\newcommand{\sube}{
\setlength{\unitlength}{0.069mm}
\raisebox{-56.1\unitlength}{
\begin{picture}(288.1,112.2)( 0.0,-56.1)
\put( 0.0, 0.0){\line(1,0){88.0}}
\bezier{\densityw}(88.0, 0.0)(88.0,-4.0)(88.6,-8.0)
\bezier{\densityw}(88.6,-8.0)(89.1,-11.9)(90.3,-15.8)
\bezier{\densityw}(90.3,-15.8)(91.4,-19.6)(93.1,-23.3)
\bezier{\densityw}(93.1,-23.3)(94.7,-26.9)(96.9,-30.3)
\bezier{\densityw}(96.9,-30.3)(99.1,-33.7)(101.7,-36.7)
\bezier{\densityw}(101.7,-36.7)(104.3,-39.7)(107.4,-42.4)
\bezier{\densityw}(107.4,-42.4)(110.4,-45.0)(113.8,-47.2)
\bezier{\densityw}(113.8,-47.2)(117.1,-49.3)(120.8,-51.0)
\bezier{\densityw}(120.8,-51.0)(124.4,-52.7)(128.3,-53.8)
\bezier{\densityw}(128.3,-53.8)(132.1,-54.9)(136.1,-55.5)
\bezier{\densityw}(136.1,-55.5)(140.1,-56.1)(144.1,-56.1)
\bezier{\densityw}(144.1,-56.1)(148.1,-56.1)(152.0,-55.5)
\bezier{\densityw}(152.0,-55.5)(156.0,-54.9)(159.9,-53.8)
\bezier{\densityw}(159.9,-53.8)(163.7,-52.7)(167.4,-51.0)
\bezier{\densityw}(167.4,-51.0)(171.0,-49.3)(174.4,-47.2)
\bezier{\densityw}(174.4,-47.2)(177.8,-45.0)(180.8,-42.4)
\bezier{\densityw}(180.8,-42.4)(183.8,-39.7)(186.4,-36.7)
\bezier{\densityw}(186.4,-36.7)(189.1,-33.7)(191.2,-30.3)
\bezier{\densityw}(191.2,-30.3)(193.4,-26.9)(195.1,-23.3)
\bezier{\densityw}(195.1,-23.3)(196.7,-19.6)(197.9,-15.8)
\bezier{\densityw}(197.9,-15.8)(199.0,-11.9)(199.6,-8.0)
\bezier{\densityw}(199.6,-8.0)(200.1,-4.0)(200.1, 0.0)
\bezier{\densityw}(200.1, 0.0)(200.1, 0.4)(200.1, 0.9)
\bezier{\densityw}(199.4, 8.8)(199.4, 9.2)(199.3, 9.6)
\bezier{\densityw}(197.4,17.3)(197.3,17.7)(197.1,18.2)
\bezier{\densityw}(194.0,25.5)(193.8,25.9)(193.6,26.2)
\bezier{\densityw}(189.4,33.0)(189.2,33.3)(188.9,33.7)
\bezier{\densityw}(183.7,39.7)(183.4,40.0)(183.1,40.3)
\bezier{\densityw}(177.0,45.4)(176.7,45.6)(176.3,45.9)
\bezier{\densityw}(169.5,50.0)(169.1,50.2)(168.7,50.4)
\bezier{\densityw}(161.4,53.3)(161.0,53.5)(160.5,53.6)
\bezier{\densityw}(152.8,55.4)(152.4,55.5)(152.0,55.5)
\put(200.1, 0.0){\line(1,0){88.0}}
\bezier{\densityw}(144.1,56.1)(140.0,56.1)(136.1,55.5)
\bezier{\densityw}(136.1,55.5)(132.1,54.9)(128.3,53.8)
\bezier{\densityw}(128.3,53.8)(124.4,52.7)(120.8,51.0)
\bezier{\densityw}(120.8,51.0)(117.1,49.3)(113.7,47.2)
\bezier{\densityw}(113.7,47.2)(110.4,45.0)(107.3,42.4)
\bezier{\densityw}(107.3,42.4)(104.3,39.8)(101.7,36.7)
\bezier{\densityw}(101.7,36.7)(99.1,33.7)(96.9,30.3)
\bezier{\densityw}(96.9,30.3)(94.7,27.0)(93.1,23.3)
\bezier{\densityw}(93.1,23.3)(91.4,19.7)(90.3,15.8)
\bezier{\densityw}(90.3,15.8)(89.1,12.0)(88.6, 8.0)
\bezier{\densityw}(88.6, 8.0)(88.0, 4.0)(88.0, 0.0)
\put(144.1,56.1){\line(0,-1){112.2}}
\end{picture}
} }
\newcommand{\subf}{
\setlength{\unitlength}{0.069mm}
\raisebox{-56.1\unitlength}{
\begin{picture}(288.1,112.1)( 0.0,-56.1)
\put( 0.0, 0.0){\line(1,0){88.0}}
\bezier{\densityw}(88.0, 0.0)(88.0,-4.0)(88.6,-8.0)
\bezier{\densityw}(88.6,-8.0)(89.1,-11.9)(90.3,-15.8)
\bezier{\densityw}(90.3,-15.8)(91.4,-19.6)(93.1,-23.3)
\bezier{\densityw}(93.1,-23.3)(94.7,-26.9)(96.9,-30.3)
\bezier{\densityw}(96.9,-30.3)(99.1,-33.7)(101.7,-36.7)
\bezier{\densityw}(101.7,-36.7)(104.3,-39.7)(107.4,-42.4)
\bezier{\densityw}(107.4,-42.4)(110.4,-45.0)(113.8,-47.2)
\bezier{\densityw}(113.8,-47.2)(117.1,-49.3)(120.8,-51.0)
\bezier{\densityw}(120.8,-51.0)(124.4,-52.7)(128.3,-53.8)
\bezier{\densityw}(128.3,-53.8)(132.1,-54.9)(136.1,-55.5)
\bezier{\densityw}(136.1,-55.5)(140.1,-56.1)(144.1,-56.1)
\bezier{\densityw}(144.1,-56.1)(148.1,-56.1)(152.0,-55.5)
\bezier{\densityw}(152.0,-55.5)(156.0,-54.9)(159.9,-53.8)
\bezier{\densityw}(159.9,-53.8)(163.7,-52.7)(167.4,-51.0)
\bezier{\densityw}(167.4,-51.0)(171.0,-49.3)(174.4,-47.2)
\bezier{\densityw}(174.4,-47.2)(177.8,-45.0)(180.8,-42.4)
\bezier{\densityw}(180.8,-42.4)(183.8,-39.7)(186.4,-36.7)
\bezier{\densityw}(186.4,-36.7)(189.1,-33.7)(191.2,-30.3)
\bezier{\densityw}(191.2,-30.3)(193.4,-26.9)(195.1,-23.3)
\bezier{\densityw}(195.1,-23.3)(196.7,-19.6)(197.9,-15.8)
\bezier{\densityw}(197.9,-15.8)(199.0,-11.9)(199.6,-8.0)
\bezier{\densityw}(199.6,-8.0)(200.1,-4.0)(200.1, 0.0)
\bezier{\densityw}(200.1, 0.0)(200.1, 4.0)(199.6, 8.0)
\bezier{\densityw}(199.6, 8.0)(199.0,11.9)(197.9,15.8)
\bezier{\densityw}(197.9,15.8)(196.7,19.6)(195.1,23.3)
\bezier{\densityw}(195.1,23.3)(193.4,26.9)(191.2,30.3)
\bezier{\densityw}(191.2,30.3)(189.1,33.7)(186.4,36.7)
\bezier{\densityw}(186.4,36.7)(183.8,39.7)(180.8,42.4)
\bezier{\densityw}(180.8,42.4)(177.8,45.0)(174.4,47.2)
\bezier{\densityw}(174.4,47.2)(171.0,49.3)(167.4,51.0)
\bezier{\densityw}(167.4,51.0)(163.7,52.7)(159.9,53.8)
\bezier{\densityw}(159.9,53.8)(156.0,54.9)(152.0,55.5)
\bezier{\densityw}(152.0,55.5)(148.1,56.1)(144.1,56.1)
\put(200.1, 0.0){\line(1,0){88.0}}
\bezier{\densityw}(144.1,56.1)(140.1,56.1)(136.1,55.5)
\bezier{\densityw}(136.1,55.5)(132.1,54.9)(128.3,53.8)
\bezier{\densityw}(128.3,53.8)(124.4,52.7)(120.8,51.0)
\bezier{\densityw}(120.8,51.0)(117.1,49.3)(113.8,47.2)
\bezier{\densityw}(113.8,47.2)(110.4,45.0)(107.4,42.4)
\bezier{\densityw}(107.4,42.4)(104.3,39.7)(101.7,36.7)
\bezier{\densityw}(101.7,36.7)(99.1,33.7)(96.9,30.3)
\bezier{\densityw}(96.9,30.3)(94.7,26.9)(93.1,23.3)
\bezier{\densityw}(93.1,23.3)(91.4,19.6)(90.3,15.8)
\bezier{\densityw}(90.3,15.8)(89.1,11.9)(88.6, 8.0)
\bezier{\densityw}(88.6, 8.0)(88.0, 4.0)(88.0, 0.0)
\multiput(144.1,56.1)(-0.00,-8.63){13}{\line(0,-1){ 0.9}}
\end{picture}
} }
\newcommand{\subg}{
\setlength{\unitlength}{0.069mm}
\raisebox{-56.1\unitlength}{
\begin{picture}(288.2,112.2)( 0.0,-56.1)
\put( 0.0, 0.0){\line(1,0){88.0}}
\bezier{\densityw}(88.0, 0.0)(88.0,-0.4)(88.0,-0.9)
\bezier{\densityw}(88.7,-8.8)(88.8,-9.2)(88.8,-9.6)
\bezier{\densityw}(90.7,-17.3)(90.9,-17.7)(91.0,-18.2)
\bezier{\densityw}(94.1,-25.5)(94.3,-25.9)(94.5,-26.2)
\bezier{\densityw}(98.7,-33.0)(99.0,-33.3)(99.2,-33.7)
\bezier{\densityw}(104.4,-39.7)(104.7,-40.0)(105.1,-40.3)
\bezier{\densityw}(111.1,-45.4)(111.5,-45.6)(111.8,-45.9)
\bezier{\densityw}(118.6,-50.0)(119.0,-50.2)(119.4,-50.4)
\bezier{\densityw}(126.8,-53.3)(127.2,-53.5)(127.6,-53.6)
\bezier{\densityw}(135.3,-55.4)(135.7,-55.5)(136.2,-55.5)
\bezier{\densityw}(144.1,-56.1)(148.1,-56.1)(152.1,-55.5)
\bezier{\densityw}(152.1,-55.5)(156.0,-54.9)(159.9,-53.8)
\bezier{\densityw}(159.9,-53.8)(163.7,-52.7)(167.4,-51.0)
\bezier{\densityw}(167.4,-51.0)(171.0,-49.3)(174.4,-47.2)
\bezier{\densityw}(174.4,-47.2)(177.8,-45.0)(180.8,-42.4)
\bezier{\densityw}(180.8,-42.4)(183.8,-39.8)(186.5,-36.7)
\bezier{\densityw}(186.5,-36.7)(189.1,-33.7)(191.2,-30.3)
\bezier{\densityw}(191.2,-30.3)(193.4,-27.0)(195.1,-23.3)
\bezier{\densityw}(195.1,-23.3)(196.7,-19.7)(197.9,-15.8)
\bezier{\densityw}(197.9,-15.8)(199.0,-12.0)(199.6,-8.0)
\bezier{\densityw}(199.6,-8.0)(200.2,-4.0)(200.2,-0.0)
\bezier{\densityw}(200.2,-0.0)(200.2, 0.4)(200.1, 0.9)
\bezier{\densityw}(199.5, 8.8)(199.4, 9.2)(199.3, 9.6)
\bezier{\densityw}(197.4,17.3)(197.3,17.7)(197.1,18.2)
\bezier{\densityw}(194.0,25.4)(193.8,25.8)(193.6,26.2)
\bezier{\densityw}(189.4,33.0)(189.2,33.3)(188.9,33.7)
\bezier{\densityw}(183.7,39.6)(183.4,40.0)(183.1,40.3)
\bezier{\densityw}(177.0,45.4)(176.7,45.6)(176.3,45.9)
\bezier{\densityw}(169.5,50.0)(169.1,50.2)(168.7,50.4)
\bezier{\densityw}(161.4,53.3)(161.0,53.5)(160.6,53.6)
\bezier{\densityw}(152.8,55.4)(152.4,55.4)(152.0,55.5)
\put(200.2,-0.0){\line(1,0){88.0}}
\bezier{\densityw}(144.1,56.1)(140.1,56.1)(136.1,55.5)
\bezier{\densityw}(136.1,55.5)(132.1,54.9)(128.3,53.8)
\bezier{\densityw}(128.3,53.8)(124.4,52.7)(120.8,51.0)
\bezier{\densityw}(120.8,51.0)(117.1,49.3)(113.8,47.2)
\bezier{\densityw}(113.8,47.2)(110.4,45.0)(107.4,42.4)
\bezier{\densityw}(107.4,42.4)(104.3,39.7)(101.7,36.7)
\bezier{\densityw}(101.7,36.7)(99.1,33.7)(96.9,30.3)
\bezier{\densityw}(96.9,30.3)(94.7,26.9)(93.1,23.3)
\bezier{\densityw}(93.1,23.3)(91.4,19.6)(90.3,15.8)
\bezier{\densityw}(90.3,15.8)(89.1,11.9)(88.6, 8.0)
\bezier{\densityw}(88.6, 8.0)(88.0, 4.0)(88.0, 0.0)
\put(144.1,56.1){\line(0,-1){112.2}}
\end{picture}
} }
\newcommand{\subh}{
\setlength{\unitlength}{0.069mm}
\raisebox{-56.1\unitlength}{
\begin{picture}(288.2,112.2)( 0.0,-56.1)
\put( 0.0, 0.0){\line(1,0){88.0}}
\bezier{\densityw}(88.0, 0.0)(88.0,-4.0)(88.6,-8.0)
\bezier{\densityw}(88.6,-8.0)(89.1,-11.9)(90.3,-15.8)
\bezier{\densityw}(90.3,-15.8)(91.4,-19.6)(93.1,-23.3)
\bezier{\densityw}(93.1,-23.3)(94.7,-26.9)(96.9,-30.3)
\bezier{\densityw}(96.9,-30.3)(99.1,-33.7)(101.7,-36.7)
\bezier{\densityw}(101.7,-36.7)(104.3,-39.7)(107.4,-42.4)
\bezier{\densityw}(107.4,-42.4)(110.4,-45.0)(113.8,-47.2)
\bezier{\densityw}(113.8,-47.2)(117.1,-49.3)(120.8,-51.0)
\bezier{\densityw}(120.8,-51.0)(124.4,-52.7)(128.3,-53.8)
\bezier{\densityw}(128.3,-53.8)(132.1,-54.9)(136.1,-55.5)
\bezier{\densityw}(136.1,-55.5)(140.1,-56.1)(144.1,-56.1)
\bezier{\densityw}(144.1,-56.1)(144.5,-56.1)(145.0,-56.1)
\bezier{\densityw}(152.8,-55.4)(153.3,-55.3)(153.7,-55.2)
\bezier{\densityw}(161.4,-53.3)(161.8,-53.2)(162.2,-53.0)
\bezier{\densityw}(169.5,-50.0)(169.9,-49.8)(170.3,-49.6)
\bezier{\densityw}(177.0,-45.4)(177.4,-45.1)(177.7,-44.8)
\bezier{\densityw}(183.7,-39.6)(184.0,-39.3)(184.3,-39.0)
\bezier{\densityw}(189.4,-33.0)(189.7,-32.6)(190.0,-32.2)
\bezier{\densityw}(194.0,-25.4)(194.2,-25.1)(194.4,-24.7)
\bezier{\densityw}(197.4,-17.3)(197.5,-16.9)(197.7,-16.5)
\bezier{\densityw}(199.5,-8.8)(199.5,-8.3)(199.6,-7.9)
\bezier{\densityw}(200.2, 0.0)(200.2, 4.0)(199.6, 8.0)
\bezier{\densityw}(199.6, 8.0)(199.0,12.0)(197.9,15.8)
\bezier{\densityw}(197.9,15.8)(196.7,19.7)(195.1,23.3)
\bezier{\densityw}(195.1,23.3)(193.4,27.0)(191.2,30.3)
\bezier{\densityw}(191.2,30.3)(189.1,33.7)(186.5,36.7)
\bezier{\densityw}(186.5,36.7)(183.8,39.8)(180.8,42.4)
\bezier{\densityw}(180.8,42.4)(177.8,45.0)(174.4,47.2)
\bezier{\densityw}(174.4,47.2)(171.0,49.3)(167.4,51.0)
\bezier{\densityw}(167.4,51.0)(163.7,52.7)(159.9,53.8)
\bezier{\densityw}(159.9,53.8)(156.0,54.9)(152.1,55.5)
\bezier{\densityw}(152.1,55.5)(148.1,56.1)(144.1,56.1)
\put(200.2, 0.0){\line(1,0){88.0}}
\bezier{\densityw}(144.1,56.1)(143.6,56.1)(143.2,56.1)
\bezier{\densityw}(135.3,55.4)(134.9,55.3)(134.4,55.2)
\bezier{\densityw}(126.8,53.3)(126.3,53.2)(125.9,53.1)
\bezier{\densityw}(118.6,50.0)(118.2,49.8)(117.8,49.6)
\bezier{\densityw}(111.1,45.4)(110.8,45.1)(110.4,44.8)
\bezier{\densityw}(104.4,39.7)(104.1,39.3)(103.8,39.0)
\bezier{\densityw}(98.7,33.0)(98.5,32.6)(98.2,32.2)
\bezier{\densityw}(94.1,25.5)(93.9,25.1)(93.7,24.7)
\bezier{\densityw}(90.7,17.3)(90.6,16.9)(90.5,16.5)
\bezier{\densityw}(88.7, 8.8)(88.6, 8.3)(88.6, 7.9)
\put(144.1,56.1){\line(0,-1){112.2}}
\end{picture}
} }
\newcommand{\subi}{
\setlength{\unitlength}{0.069mm}
\raisebox{-56.1\unitlength}{
\begin{picture}(288.2,112.2)( 0.0,-56.1)
\put( 0.0, 0.0){\line(1,0){88.0}}
\bezier{\densityw}(88.0, 0.0)(88.0,-0.4)(88.0,-0.9)
\bezier{\densityw}(88.7,-8.8)(88.8,-9.2)(88.8,-9.6)
\bezier{\densityw}(90.7,-17.3)(90.9,-17.7)(91.0,-18.2)
\bezier{\densityw}(94.1,-25.5)(94.3,-25.9)(94.5,-26.2)
\bezier{\densityw}(98.7,-33.0)(99.0,-33.3)(99.2,-33.7)
\bezier{\densityw}(104.4,-39.7)(104.7,-40.0)(105.1,-40.3)
\bezier{\densityw}(111.1,-45.4)(111.5,-45.6)(111.8,-45.9)
\bezier{\densityw}(118.6,-50.0)(119.0,-50.2)(119.4,-50.4)
\bezier{\densityw}(126.8,-53.3)(127.2,-53.5)(127.6,-53.6)
\bezier{\densityw}(135.3,-55.4)(135.7,-55.5)(136.2,-55.5)
\bezier{\densityw}(144.1,-56.1)(144.5,-56.1)(145.0,-56.1)
\bezier{\densityw}(152.9,-55.4)(153.3,-55.3)(153.7,-55.2)
\bezier{\densityw}(161.4,-53.3)(161.8,-53.2)(162.2,-53.1)
\bezier{\densityw}(169.5,-50.0)(169.9,-49.8)(170.3,-49.6)
\bezier{\densityw}(177.0,-45.4)(177.4,-45.1)(177.8,-44.8)
\bezier{\densityw}(183.7,-39.7)(184.0,-39.3)(184.4,-39.0)
\bezier{\densityw}(189.5,-33.0)(189.7,-32.6)(190.0,-32.2)
\bezier{\densityw}(194.0,-25.5)(194.2,-25.1)(194.4,-24.7)
\bezier{\densityw}(197.4,-17.3)(197.6,-16.9)(197.7,-16.5)
\bezier{\densityw}(199.5,-8.8)(199.5,-8.3)(199.6,-7.9)
\bezier{\densityw}(200.2, 0.0)(200.2, 4.0)(199.6, 8.0)
\bezier{\densityw}(199.6, 8.0)(199.0,11.9)(197.9,15.8)
\bezier{\densityw}(197.9,15.8)(196.8,19.6)(195.1,23.3)
\bezier{\densityw}(195.1,23.3)(193.4,26.9)(191.3,30.3)
\bezier{\densityw}(191.3,30.3)(189.1,33.7)(186.5,36.7)
\bezier{\densityw}(186.5,36.7)(183.8,39.7)(180.8,42.4)
\bezier{\densityw}(180.8,42.4)(177.8,45.0)(174.4,47.2)
\bezier{\densityw}(174.4,47.2)(171.0,49.3)(167.4,51.0)
\bezier{\densityw}(167.4,51.0)(163.7,52.7)(159.9,53.8)
\bezier{\densityw}(159.9,53.8)(156.0,54.9)(152.1,55.5)
\bezier{\densityw}(152.1,55.5)(148.1,56.1)(144.1,56.1)
\put(200.2, 0.0){\line(1,0){88.0}}
\bezier{\densityw}(144.1,56.1)(140.1,56.1)(136.1,55.5)
\bezier{\densityw}(136.1,55.5)(132.1,54.9)(128.3,53.8)
\bezier{\densityw}(128.3,53.8)(124.4,52.7)(120.8,51.0)
\bezier{\densityw}(120.8,51.0)(117.2,49.3)(113.8,47.2)
\bezier{\densityw}(113.8,47.2)(110.4,45.0)(107.4,42.4)
\bezier{\densityw}(107.4,42.4)(104.3,39.7)(101.7,36.7)
\bezier{\densityw}(101.7,36.7)(99.1,33.7)(96.9,30.3)
\bezier{\densityw}(96.9,30.3)(94.8,26.9)(93.1,23.3)
\bezier{\densityw}(93.1,23.3)(91.4,19.6)(90.3,15.8)
\bezier{\densityw}(90.3,15.8)(89.2,11.9)(88.6, 8.0)
\bezier{\densityw}(88.6, 8.0)(88.0, 4.0)(88.0, 0.0)
\multiput(144.1,56.1)(-0.00,-8.63){13}{\line(0,-1){ 0.9}}
\end{picture}
} }
\newcommand{\subj}{
\setlength{\unitlength}{0.069mm}
\raisebox{-56.1\unitlength}{
\begin{picture}(288.1,112.2)( 0.0,-56.1)
\put( 0.0, 0.0){\line(1,0){88.0}}
\bezier{\densityw}(88.0, 0.0)(88.0,-4.0)(88.6,-8.0)
\bezier{\densityw}(88.6,-8.0)(89.1,-11.9)(90.3,-15.8)
\bezier{\densityw}(90.3,-15.8)(91.4,-19.6)(93.1,-23.3)
\bezier{\densityw}(93.1,-23.3)(94.7,-26.9)(96.9,-30.3)
\bezier{\densityw}(96.9,-30.3)(99.1,-33.7)(101.7,-36.7)
\bezier{\densityw}(101.7,-36.7)(104.3,-39.7)(107.4,-42.4)
\bezier{\densityw}(107.4,-42.4)(110.4,-45.0)(113.8,-47.2)
\bezier{\densityw}(113.8,-47.2)(117.1,-49.3)(120.8,-51.0)
\bezier{\densityw}(120.8,-51.0)(124.4,-52.7)(128.3,-53.8)
\bezier{\densityw}(128.3,-53.8)(132.1,-54.9)(136.1,-55.5)
\bezier{\densityw}(136.1,-55.5)(140.1,-56.1)(144.1,-56.1)
\bezier{\densityw}(144.1,-56.1)(148.1,-56.1)(152.0,-55.5)
\bezier{\densityw}(152.0,-55.5)(156.0,-54.9)(159.9,-53.8)
\bezier{\densityw}(159.9,-53.8)(163.7,-52.7)(167.4,-51.0)
\bezier{\densityw}(167.4,-51.0)(171.0,-49.3)(174.4,-47.2)
\bezier{\densityw}(174.4,-47.2)(177.8,-45.0)(180.8,-42.4)
\bezier{\densityw}(180.8,-42.4)(183.8,-39.7)(186.4,-36.7)
\bezier{\densityw}(186.4,-36.7)(189.1,-33.7)(191.2,-30.3)
\bezier{\densityw}(191.2,-30.3)(193.4,-26.9)(195.1,-23.3)
\bezier{\densityw}(195.1,-23.3)(196.7,-19.6)(197.9,-15.8)
\bezier{\densityw}(197.9,-15.8)(199.0,-11.9)(199.6,-8.0)
\bezier{\densityw}(199.6,-8.0)(200.1,-4.0)(200.1, 0.0)
\bezier{\densityw}(200.1, 0.0)(200.1, 0.4)(200.1, 0.9)
\bezier{\densityw}(199.4, 8.8)(199.4, 9.2)(199.3, 9.6)
\bezier{\densityw}(197.4,17.3)(197.3,17.7)(197.1,18.2)
\bezier{\densityw}(194.0,25.5)(193.8,25.9)(193.6,26.2)
\bezier{\densityw}(189.4,33.0)(189.2,33.3)(188.9,33.7)
\bezier{\densityw}(183.7,39.7)(183.4,40.0)(183.1,40.3)
\bezier{\densityw}(177.0,45.4)(176.7,45.6)(176.3,45.9)
\bezier{\densityw}(169.5,50.0)(169.1,50.2)(168.7,50.4)
\bezier{\densityw}(161.4,53.3)(161.0,53.5)(160.5,53.6)
\bezier{\densityw}(152.8,55.4)(152.4,55.5)(152.0,55.5)
\put(200.1, 0.0){\line(1,0){88.0}}
\bezier{\densityw}(144.1,56.1)(143.6,56.1)(143.2,56.1)
\bezier{\densityw}(135.3,55.4)(134.9,55.3)(134.4,55.2)
\bezier{\densityw}(126.7,53.3)(126.3,53.2)(125.9,53.1)
\bezier{\densityw}(118.6,50.0)(118.2,49.8)(117.8,49.6)
\bezier{\densityw}(111.1,45.4)(110.7,45.1)(110.4,44.8)
\bezier{\densityw}(104.4,39.7)(104.1,39.3)(103.8,39.0)
\bezier{\densityw}(98.7,33.0)(98.4,32.6)(98.2,32.2)
\bezier{\densityw}(94.1,25.5)(93.9,25.1)(93.7,24.7)
\bezier{\densityw}(90.7,17.3)(90.6,16.9)(90.5,16.5)
\bezier{\densityw}(88.7, 8.8)(88.6, 8.3)(88.5, 7.9)
\multiput(144.1,56.1)(0.00,-8.63){13}{\line(0,-1){ 0.9}}
\end{picture}
} }

\begin{figure}[htb]
  \[
  \begin{array}{rcccc}
\mbox{ Type 1}:&\suba & (\gamma=\Gamma);&      &      \\[4ex]
\mbox{ Type 2}:&\subb,&\subc,           &\subd,&\sube;\\[4ex]
\mbox{ Type 3}:&\subf;&                 &      &      \\[4ex]
\mbox{ Type 4}:&\subg,&\subh;           &      &      \\[4ex]
\mbox{ Type 5}:&\subi,&\subj.           &      &
  \end{array}
  \]
  \caption{ The subgraphs $\gamma$ contributing to the large $k^2$ expansion }
  \label{subgraphs}
\end{figure}
The reduced graphs $\Gamma/\gamma$ correspond to the dotted lines and
can be obtained by shrinking all solid lines to a point.  In such a
way, we obtain that for the second and third type (see Fig.~3)
$J_{\Gamma/\gamma}$ corresponds to a massive tadpole, for the fourth type
we obtain a product of two massive tadpoles, while for the fifth type we
get a two-loop massive vacuum integral with three internal lines,
corresponding to Fig.~2. All other integrals (occurring in separate
terms of the expansion) can also be evaluated.

We have realized this algorithm also by use of the {\sf REDUCE} system
$^{14}$.
The analytical expressions for the coefficients
contain powers and logarithms of the masses and the external momentum
squared, and also the function of masses corresponding to the two-loop
vacuum diagram (Fig.~2) that can be expressed in terms of dilogarithms.
For some special cases occurring in the Standard Model, we compared
our expansion with the result of the numerical integration~$^{11}$,
and we found good agreement in the region above the highest threshold.
The asymptotic value as $k^2 \to \infty$ is, of course,
$-6 \zeta(3) \pi^4/k^2$ (in the case of unit powers of denominators).

Recently some of these algorithms have been used
for the renormalization group analysis of hadronic decays of a
charged Higgs boson~$^{17}$.
%
%
\vglue 0.6cm
{\elevenbf\noindent 2. Three- and four-point massless diagrams}
\vglue 0.4cm
The methods considered in Section~1 can be also applied to massive
diagrams with larger number of external lines. When constructing the expansion
for large values of the external momentum invariants, we shall
need results for the corresponding diagrams with massless
internal particles. The examination of such diagrams is
also important for calculation of some types of radiative corrections
and for analysis of some theoretical models.

In papers~$^{3,5}$
we considered two-loop three- and four-point
contributions with off-shell external momenta.
We have also considered three- and four-point ladder contributions with
an arbitrary number of loops~$^{4}$.
We used the following tools:
(i) the Feynman
parametric representation, (ii) the ``uniqueness'' conditions~$^{18}$,
(iii)~Mellin--Barnes contour integrals and
(iv) Fourier transform to the coordinate space.
We considered only scalar diagrams (corresponding to
the massless $\phi^3$ theory), because expressions occurring in realistic
calculations can be reduced to such scalar integrals.

\newcommand{\planar}{
\setlength {\unitlength}{0.47mm}
\centering
\begin{picture}(150,58)(0,0)
\put (30,30) {\line(1,0){20}}
\put (50,30) {\line(3,1){80}}
\put (50,30) {\line(3,-1){80}}
\put (80,20) {\line(0,1){20}}
\put (110,10) {\line(0,1){40}}
\put (50,30) {\circle*{2}}
\put (80,20) {\circle*{2}}
\put (80,40) {\circle*{2}}
\put (110,10) {\circle*{2}}
\put (110,50) {\circle*{2}}
\put (22,30) {\makebox(0,0)[bl]{\large $p_3$}}
\put (134,52) {\makebox(0,0)[bl]{\large $p_1$}}
\put (134,7) {\makebox(0,0)[bl]{\large $p_2$}}
\put (75,0){\makebox(0,0)[c]{\large ${\mbox{(a)}}$}}
\end{picture}
}

\newcommand{\nonplanar}{
\setlength {\unitlength}{0.47mm}
\centering
\begin{picture}(150,58)(0,0)
\put (30,30) {\line(1,0){20}}
\put (50,30) {\line(3,1){80}}
\put (50,30) {\line(3,-1){80}}
\put (80,20) {\line(1,1){30}}
\put (80,40) {\line(1,-1){30}}
\put (50,30) {\circle*{2}}
\put (80,20) {\circle*{2}}
\put (80,40) {\circle*{2}}
\put (110,50) {\circle*{2}}
\put (110,10) {\circle*{2}}
\put (22,30) {\makebox(0,0)[bl]{\large $p_3$}}
\put (134,52) {\makebox(0,0)[bl]{\large $p_1$}}
\put (134,7) {\makebox(0,0)[bl]{\large $p_2$}}
\put (75,0){\makebox(0,0)[c]{\large ${\mbox{(b)}}$}}
\end{picture}
}

For the two-loop three-point diagram $C^{(2)}$ (see Fig.~4a),
\begin{figure}[bth]
\[
\begin{array}{cc}
\planar & \nonplanar
\end{array}
\]
\caption{The planar (a) and non-planar (b) two-loop three-point diagrams}
\end{figure}
we obtained a simple integral representation,
\be
\label{C2}
C^{(2)}(p_1^2, p_2^2, p_3^2) =
-\frac{1}{2} \left( \frac{i \pi^2}{p_3^2} \right)^2
\Int_0^1 \frac{d \xi \; \ln{\xi}}{y \xi^2 + (1-x-y) \xi + x}
 \left( \ln{\frac{y}{x}} + \ln{\xi} \right)
 \left( \ln{\frac{y}{x}} + 2 \ln{\xi} \right)
\ee
with $x \equiv p_1^2/p_3^2, \;\; y \equiv p_2^2/p_3^2 $.
The integral (\ref{C2}) can be easily calculated in terms of
polylogarithms $\mbox{Li}_N$ $^{10}$
with $N \leq 4$.

For the non-planar (``crossed'') two-loop three-point diagram
$\widetilde{C}^{(2)}(p_1^2, p_2^2, p_3^2)$ (see Fig.~4b),
we found that
\be
\label{Cc2C1}
\vspace{-1mm}
\widetilde{C}^{(2)}(p_1^2, p_2^2, p_3^2)
= \left( C^{(1)}(p_1^2, p_2^2, p_3^2) \right) ^2 \; ,
\ee
where $C^{(1)}$ is a function corresponding to one-loop triangle diagram,
\be
\label{C1}
C^{(1)}(p_1^2, p_2^2, p_3^2) = -\frac{i \pi^2}{p_3^2} \Int_0^1
 \frac{d\xi}{y \xi^2 + (1-x-y) \xi + x}
 \left( \ln{\frac{y}{x}} + 2 \ln{\xi} \right) \; .
\ee
Since the integral (\ref{C1}) can be evaluated in terms of
dilogarithms $\mbox{Li}_2$, the result for $\widetilde{C}^{(2)}$
contains products of dilogarithms.

We examined a four-point ladder diagram $D^{(2)}$
\begin{figure}[bth]
\centering
\setlength {\unitlength}{0.7mm}
\begin{picture}(150,50)(0,0)
\put (30,10) {\line(1,0){90}}
\put (30,40) {\line(1,0){90}}
\put (45,10) {\line(0,1){30}}
\put (75,10) {\line(0,1){30}}
\put (105,10) {\line(0,1){30}}
\put (45,10) {\circle*{2}}
\put (45,40) {\circle*{2}}
\put (75,10) {\circle*{2}}
\put (75,40) {\circle*{2}}
\put (105,10) {\circle*{2}}
\put (105,40) {\circle*{2}}
\put (22,10) {\makebox(0,0)[bl]{\large $k_1$}}
\put (22,38) {\makebox(0,0)[bl]{\large $k_2$}}
\put (124,38) {\makebox(0,0)[bl]{\large $k_3$}}
\put (124,10) {\makebox(0,0)[bl]{\large $k_4$}}
\end{picture}
\caption{The ``double box'' diagram}
\end{figure}
(``double box", see Fig.~5),
and showed that the corresponding function can be reduced to $C^{(2)}$,
namely,
\be
\label{D2}
D^{(2)} (k_1^2, k_2^2, k_3^2, k_4^2, s, t)
= t \; C^{(2)} (k_1^2 \; k_3^2 ,\; k_2^2 \; k_4^2 , \; s \; t) ,
\vspace{-1mm}
\ee
where $s \equiv (k_1+k_2)^2 $ and $ t \equiv (k_2+k_3)^2$ are
usual Mandelstam variables. Therefore, the ``double box" can also
be evaluated in terms of $\mbox{Li}_N$ with $N \leq 4$.

\newcommand{\threepoint}{
\setlength {\unitlength}{0.43mm}
\begin{picture}(152,75)(0,0)
\put (0,35) {\line(1,0){16}}
\put (16,35) {\line(4,1){131}}
\put (16,35) {\line(4,-1){131}}
\put (40,29) {\line(0,1){12}}
\put (64,23) {\line(0,1){24}}
\put (112,11) {\line(0,1){48}}
\put (136,5) {\line(0,1){60}}
\put (6,38) {\vector(1,0){9}}
\put (0,43) {\makebox(0,0)[bl]{\large $p_3$}}
\put (146,70) {\vector(-4,-1){9}}
\put (149,69) {\makebox(0,0)[bl]{\large $p_1$}}
\put (146,0) {\vector(-4,1){9}}
\put (149,0) {\makebox(0,0)[bl]{\large $p_2$}}
\put (77,35) {\makebox(0,0)[bl]{\huge $. \; . \; .$}}
\put (75,0){\makebox(0,0)[c]{\large ${\mbox{(a)}}$}}
\end{picture}
}

\newcommand{\fourpoint}{
\setlength {\unitlength}{0.43mm}
\begin{picture}(152,50)(0,-12)
\put (5,10) {\line(1,0){143}}
\put (5,40) {\line(1,0){143}}
\put (18,10) {\line(0,1){30}}
\put (44,10) {\line(0,1){30}}
\put (70,10) {\line(0,1){30}}
\put (109,10) {\line(0,1){30}}
\put (135,10) {\line(0,1){30}}
\put (7,8) {\vector(1,0){8}}
\put (0,0) {\makebox(0,0)[bl]{\large $k_1$}}
\put (7,42) {\vector(1,0){8}}
\put (0,47) {\makebox(0,0)[bl]{\large $k_2$}}
\put (145,42) {\vector(-1,0){8}}
\put (149,47) {\makebox(0,0)[bl]{\large $k_3$}}
\put (145,8) {\vector(-1,0){8}}
\put (149,0) {\makebox(0,0)[bl]{\large $k_4$}}
\put (78,25) {\makebox(0,0)[bl]{\huge $.\; .\; .$}}
\put (75,0){\makebox(0,0)[c]{\large ${\mbox{(b)}}$}}
\end{picture}
}

We managed to generalize some of the results for one- and two-loop
diagrams to the case of arbitrary number of loops~$^{4}$.
We considered $L$-loop three-point ladder diagrams $C^{(L)}$ (Fig.~6a),
\begin{figure}[bth]
\[
\begin{array}{ccc}
\threepoint & \hspace{8mm} & \fourpoint
\end{array}
\]
\caption{The three-point (a) and four-point (b) $L$-loop diagrams}
\end{figure}
and we obtained the following generalization of the formulae
(\ref{C1}), (\ref{C2}):
\bea
\label{CL}
C^{(L)}(p_1^2, p_2^2, p_3^2) = - \frac{1}{L! \; (L-1)!}
\left( \frac{i \pi^2}{p_3^2} \right)^L
\hspace{5cm}
\nonumber \\
\times \Int_0^1
\frac{d \xi}{y \xi^2 + (1-x-y) \xi +x} \;
\ln^{L-1} \xi \left( \lnyx + \ln{\xi}  \right)^{L-1}
\left( \lnyx + 2\ln{\xi}  \right) .
\eea
This integral can be calculated in terms of polylogarithms $\mbox{Li}_N$
with $N \leq 2L$.

Moreover, for the four-point $L$-loop ladder diagram $D^{(L)}$
(shown in Fig.~6b) we found that
\be
\label{DL}
D^{(L)}(k_1^2, k_2^2, k_3^2, k_4^2, s, t)
= t^{L-1} \; C^{(L)} (k_1^2 k_3^2, k_2^2 k_4^2, s t) .
\ee
This formula generalizes eq.~(\ref{D2}) and holds also for
one-loop and zero-loop (tree) cases. Thus, the four-point
$L$-loop diagram can be also expressed in terms of $\mbox{Li}_N$
with $N \leq 2L$. Note that another derivation of the formula
(\ref{DL}) was presented in ref.~$^{19}$.

We have checked the results for $L$-loop diagrams by reducing them
to the $(L+1)$-loop two-point functions (by means of additional
integration), and we found complete agreement with known
results for such two-point diagrams~$^{20}$.
\vglue 0.5cm
{\elevenbf \noindent Acknowledgements \hfil}
\vglue 0.4cm
I would like to thank the AIHENP-93 Organizing Committee
for their help, and the International
Science Foundation (New York) for financial support.
Research was supported, in part, by the Research Council of Norway.
%
%
\vglue 0.5cm
{\elevenbf\noindent References \hfil}
\vglue 0.4cm


\begin{thebibliography}{99}
\bibitem{DT} A. I. Davydychev and J. B. Tausk, {\elevenit Nucl. Phys.}
  {\elevenbf B397} (1993) 123.
\bibitem{DST} A. I. Davydychev, V. A. Smirnov and J. B. Tausk,
  Leiden preprint INLO-PUB--5/93 (1993);
  {\elevenit Nucl. Phys.} {\elevenbf B}, to appear.
\bibitem{UD1} N. I. Ussyukina and A. I. Davydychev, {\elevenit Phys. Lett.}
  {\elevenbf B298} (1993) 363.
\bibitem{UD2} N. I. Ussyukina and A. I. Davydychev, {\elevenit Phys. Lett.}
  {\elevenbf B305} (1993) 136.
\bibitem{UD3} N. I. Ussyukina and A. I. Davydychev, {\elevenit Yad. Fiz.}
  {\elevenbf 56} (No.~11) (1993) 172.
\bibitem{WSB} G. Weiglein, R. Scharf and M. B\"ohm,
  W\"urzburg preprint (1993).
\bibitem{dimreg} G. 'tHooft and M. Veltman, {\elevenit Nucl. Phys.}
  {\elevenbf B44} (1972) 189; \\
  C. G.~Bollini and J. J.~Giambiagi, {\elevenit Nuovo Cim.}
  {\elevenbf 12B} (1972) 20.
\bibitem{QED} G.~K\"{a}ll\'{e}n and A.~Sabry, {\elevenit Dan. Mat. Fys. Medd.}
        {\elevenbf 29} (1955) No.17;
  \\    J.~Schwinger, {\elevenit Particles, Sources and Fields}, Vol.2
        (Addison-Wesley, 1973);
  \\    D. J.~Broadhurst, {\elevenit Phys. Lett.} {\elevenbf B101} (1981) 423;
         {\elevenit Z. Phys.} {\elevenbf C47} (1990) 115;
  \\    T. H.~Chang, K. J. F.~Gaemers and W. L.~van~Neerven,
         {\elevenit Nucl. Phys.} {\elevenbf B202} (1982) 407;
  \\    A.~Djouadi, {\elevenit Nuovo Cim.} {\elevenbf 100A} (1988) 357;
  \\    B. A.~Kniehl, {\elevenit Nucl. Phys.} {\elevenbf  B347} (1990) 86;
  \\    D. J.~Broadhurst, J. Fleischer and O. V.~Tarasov,
          {\elevenit Z.Phys.} {\elevenbf C60} (1993) 287.
\bibitem{ST} R. Scharf and  J. B. Tausk,
  Leiden preprint INLO-PUB--7/93 (1993);
  {\elevenit Nucl. Phys.} {\elevenbf B}, to appear.
\bibitem{Lewin} L. Lewin, {\elevenit Polylogarithms and Accociated Functions}
  (North-Holland, Amsterdam, 1981).
\bibitem{Kre} D. Kreimer, {\elevenit Phys. Lett.} {\elevenbf B273} (1991) 277.
\bibitem{ibp}  F. V.~Tkachov,   {\elevenit Phys. Lett.}
    {\elevenbf B100} (1981) 65;
  \\ K. G.~Chetyrkin and F. V.~Tkachov,   {\elevenit Nucl.Phys.}
    {\elevenbf B192} (1981) 159.
\bibitem{jpa} A. I. Davydychev, {\elevenit J. Phys.} {\elevenbf A25} 5587.
\bibitem{reduce} A. C.~Hearn,  {\elevenit REDUCE User's Manual},
   RAND publication CP78 (Santa Monica, 1987).
\bibitem{asex}
      F. V.~Tkachov, Moscow preprint INR P-358 (1984);
  \\  G. B.~Pivovarov and F. V.~Tkachov, Moscow preprints INR P-0370, $\Pi$-459
       (1984);
  \\  K. G.~Chetyrkin and V. A.~Smirnov, Moscow preprint INR G-518 (1987);
  \\  K. G.~Chetyrkin, {\elevenit Teor. Mat. Fiz.}
         {\elevenbf 75} (1988) 26; {\elevenbf 76} (1988) 207;
         Munich preprint MPI-PAE/PTh 13/91 (1991);
  \\  S. G.~Gorishny, {\elevenit Nucl. Phys.} {\elevenbf B319} (1989) 633.
  \\  V. A.~Smirnov, {\elevenit Commun. Math. Phys.} {\elevenbf 134} (1990)
109.
\bibitem{Smi-book}
      V. A.~Smirnov, {\elevenit Renormalization and Asymptotic Expansions}
      (Birkh\"{a}user, \\ Basel, 1991).
\bibitem{CD} A. Czarnecki and A. I. Davydychev,
  Edmonton preprint Alberta--Thy-38-93 (1993).
\bibitem{uniq} A. N. Vasil'ev, Yu. M. Pis'mak and Yu. R. Honkonen,
    {\elevenit Teor. Mat. Fiz.} {\elevenbf 47} (1981) 291; \\
  N. I. Ussyukina, {\elevenit Teor. Mat. Fiz.} {\elevenbf 54}
    (1983) 124; \\
  D. I. Kazakov, {\elevenit Phys. Lett.} {\elevenbf B133} (1983) 406; \\
  D. I. Kazakov and A. V. Kotikov, {\elevenit Teor. Mat. Fiz.} {\elevenbf 73}
    (1987) 348.
\bibitem{Bro'93} D. J. Broadhurst, {\elevenit Phys. Lett.}
  {\elevenbf B307} (1993) 132.
\bibitem{BU+Bro} V. V. Belokurov and N. I. Ussyukina, {\elevenit J. Phys.}
   {\elevenbf A16} (1983) 2811;\\
  D. J. Broadhurst, {\elevenit Phys. Lett.} {\elevenbf B164} (1985) 356.
\end{thebibliography}
\end{document}